\documentclass[a4paper]{jacow}

\ifboolexpr{bool{xetex} or bool{luatex}}
 {}
 {\usepackage[utf8]{inputenc}}

\usepackage{subfig}
\usepackage{amsmath}
\usepackage{graphics}
\usepackage{amssymb}
 \usepackage{graphicx}
\usepackage{marginnote}
\usepackage{float}
\usepackage{lipsum}

\ifboolexpr{bool{jacowbiblatex}}
 {%
  \addbibresource{jacow-test.bib}
  \addbibresource{biblatex-examples.bib}
 }{}

\copyrightspace[2cm] 

\begin{document}

\title{Design Studies and Commissioning Plans for PARS Experimental Program}

\author{O. Mete\thanks{oznur.mete@manchester.ac.uk}, K. Hanahoe, G. Xia,
The University of Manchester, Manchester, UK and \\ The Cockcroft Institute, Sci-Tech Daresbury, Warrington, UK\\ M. Dover, M. Wigram, J. Wright,  J. Zhang,The University of Manchester, Manchester, UK \\ J. Smith, Tech-X UK Ltd, Sci-Tech Daresbury, Warrington, UK}

\maketitle
\begin{abstract}
PARS (Plasma Acceleration Research Station) is an electron beam driven plasma wakefield acceleration test stand proposed for VELA/CLARA facility in Daresbury Laboratory. In order to optimise various operational configurations, 2D numerical studies were performed by using VSIM for a range of parameters such as bunch length, radius, plasma density and positioning of the bunches with respect to each other for the two-beam acceleration scheme. In this paper, some of these numerical studies and considered measurement methods are presented.  
\end{abstract}
\section{Introduction} 
\begin{figure*}[htb!] 
\centering
\includegraphics[width=0.9\textwidth] {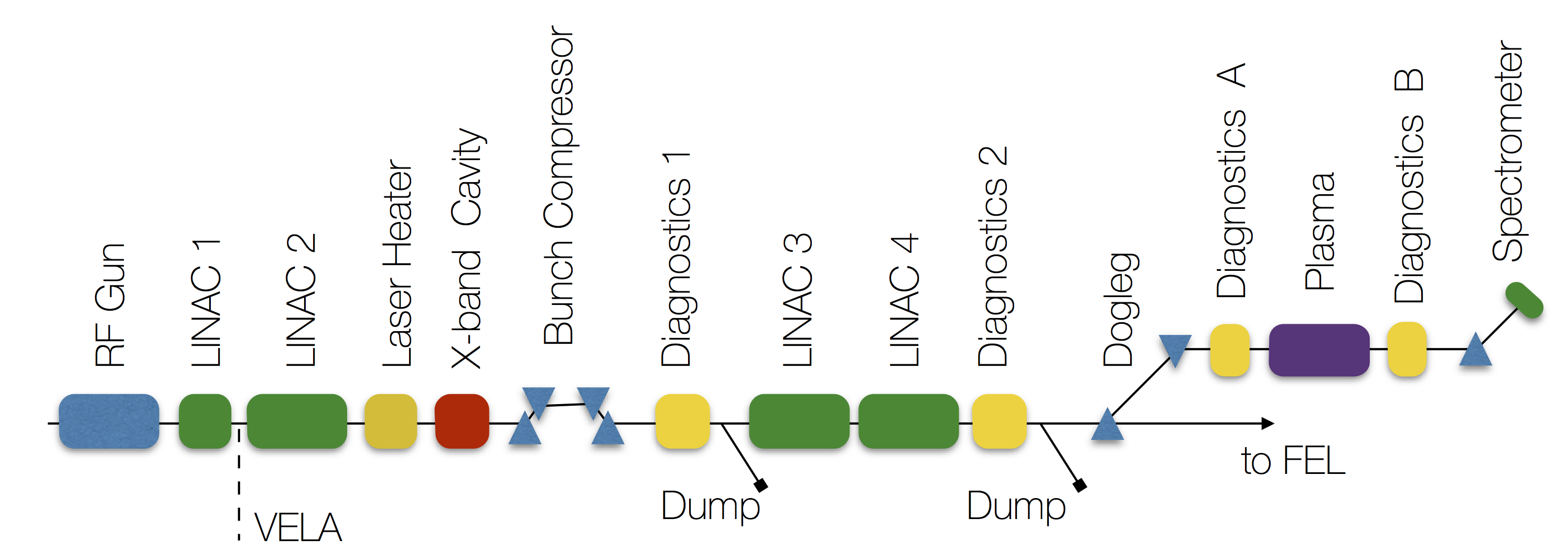} 
\caption{Layout of the CLARA beamline and PARS experimental station.}
\label{fig:layout}
\vspace{-1.0em}
\end{figure*}

PARS experimental station is planned on the soon-to-be-built VELA/CLARA beam line in the Daresbury Laboratories as shown in Fig.\ref{fig:layout} \cite{Clara, Xia}. PARS will receive an  $250\,$MeV electron beam with a flexible parameter range. This will allow the station to conduct wide range systematic studies on electron driven plasma wakefield acceleration. Program aims to explore single and two-beam operation. The former aims to study maximum achievable accelerating gradient and head-to-tail acceleration with a single electron bunch. Whereas the latter aims to demonstrate the acceleration of a witness or trailing bunch. Numerical studies reported in this paper were performed by using VSim \cite{VSim}.
\section{Single bunch acceleration} 
\begin{figure}[htb!] 
\centering
\includegraphics[width=0.45\textwidth] {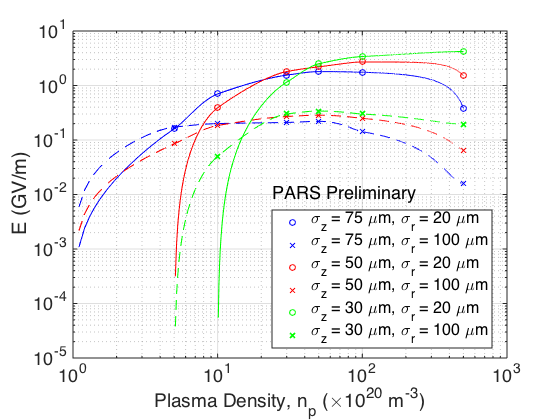} 
\caption{Accelerating gradient of the first bucket for different beam parameters as a function of the plasma density after $4.5\,$mm propagation.}
\label{fig:density_scan}
\vspace{-1.0em}
\end{figure}

The maximum achievable accelerating gradient was studied for different bunch length and radius values between $30-75\,\mu$m and $20-100\,\mu$m, respectively, considering a $250\,$MeV electron bunch with a charge of $250\,$pC. Higher wakefields of $1-3\,$GV/m for a tightly focused drive beam ($20\mu$m), and $200-300\,$MV/m for more realistic beam sizes are possible within the bunch length range between $30-75\,\mu$m (Fig.\ref{fig:density_scan}). The achieved field gradient is proportional to the plasma density and after a certain density it scales inversely proportional to the squared bunch length as predicted by the linear theory Eq.\ref{eqn:linear_theory}, 
\begin{equation}
E=240(MV/m)\bigg(\frac{N}{4\times10^{10}}\bigg)\bigg(\frac{0.6}{\sigma_z(mm)}\bigg)^2
\label{eqn:linear_theory}
\end{equation}
where N is the density of background plasma electrons and $\sigma_z$ is the bunch length.  
A realistic case of $50\,\mu$m bunch length and $100\,\mu$m bunch radius which yields $300\,$MV/m gradient for single bunch propagated $4.5\,$mm at a plasma density of $5\times10^{21}\,m^{-3}$ was selected for further studies in this paper.
\section{Two-bunch acceleration} 
\begin{figure}[htb!] 
\centering
\includegraphics[width=0.45\textwidth] {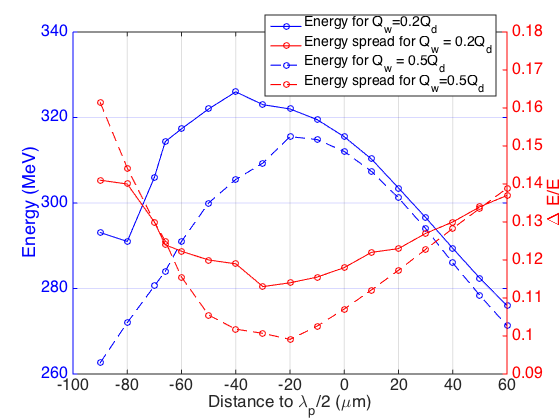}
\caption{Energy and the energy spread of the trailing bunch as a function of its distance to the driver bunch around the initial location at $\lambda_p/2$.}
\label{fig:energy_distance}
\vspace{-1.0em}
\end{figure}

\begin{figure}[htb!] 
\centering
\subfloat[]{\includegraphics[width=0.45\textwidth] {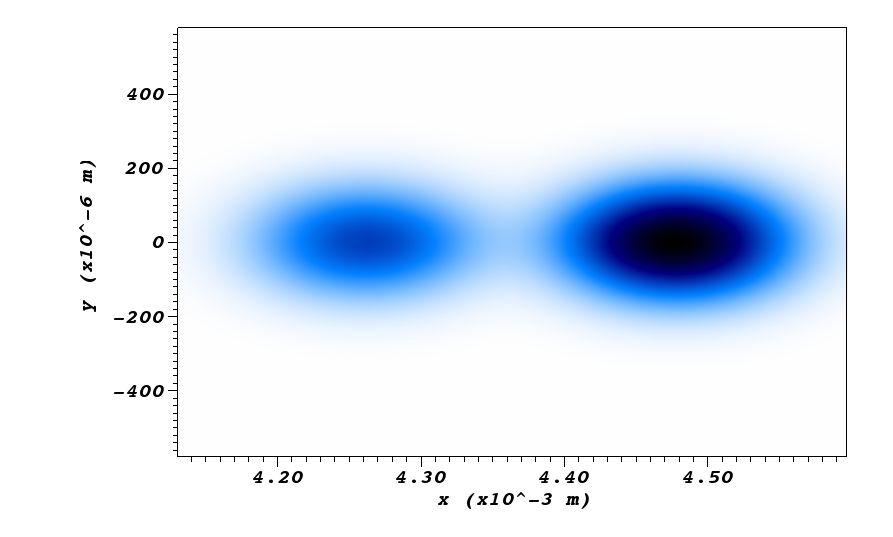}} \\
\subfloat[]{\includegraphics[width=0.45\textwidth] {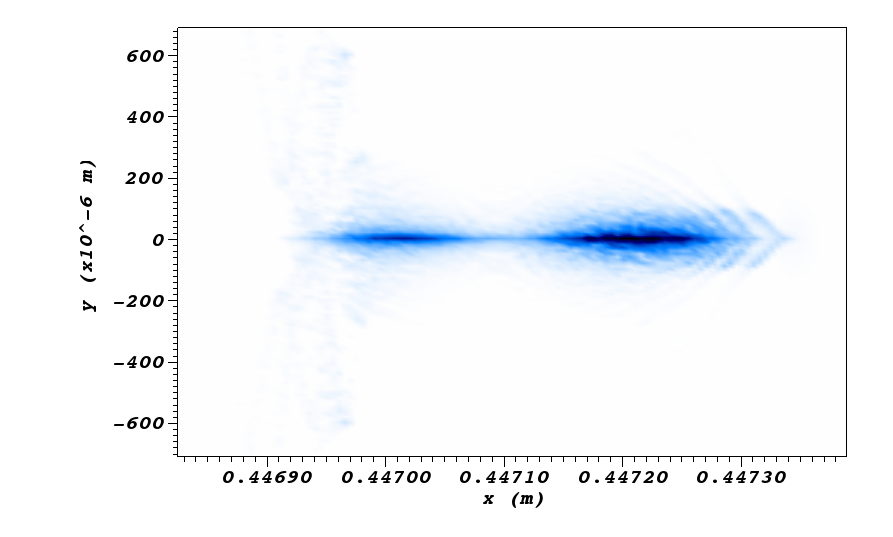}} 
\caption{(a) Initial and (b) final beam intensity distributions for the travel through $~0.5\,$m long plasma column.}
\label{fig:beam_profiles}
\vspace{-1.5em}
\end{figure}

\begin{figure}[htb!] 
\centering
\includegraphics[width=0.45\textwidth] {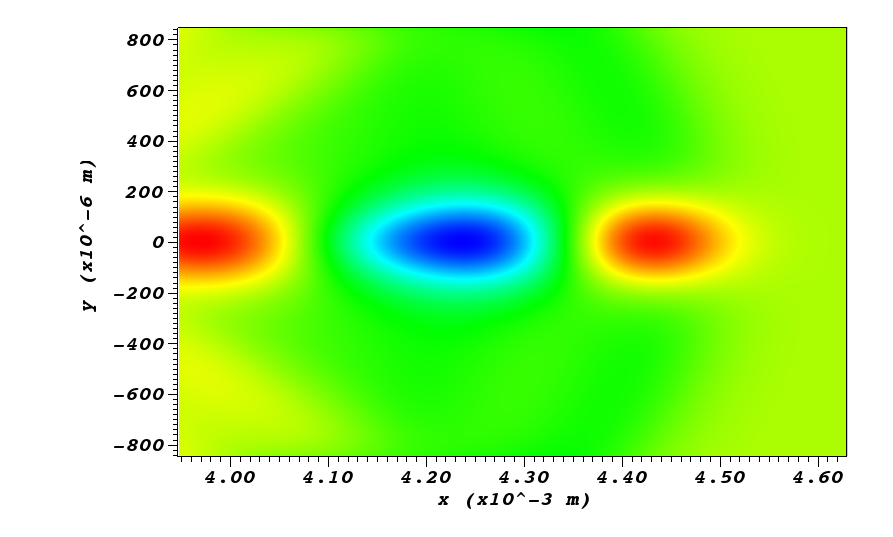}
\caption{2D field strength distribution for the longitudinal component of the plasma wakefield.}
\label{fig:2D_field}
\vspace{-1.0em}
\end{figure}

A two-beam scenario was simulated using the baseline case detailed above. A second bunch of same sizes but with a certain fraction of the drive bunch charge was initially placed half a plasma wavelength ($\lambda_p/2$) behind the centre of the driver bunch. The maximum energy gain and the minimum energy spread of the trailing bunch was found to occur between $(\lambda_p/2-40\,)\mu$m and  $(\lambda_p/2-20\,)\mu$m behind the driver bunch as shown in Fig.\ref{fig:energy_distance}. 
The initial two beam configuration shown in Fig.\ref{fig:beam_profiles}-(a) was tracked along a $~0.5\,$m long plasma column. Fig.\ref{fig:beam_profiles}-(b) shows the beam profiles evolved after $0.45\,$m reaching an energy of $315\,MeV$ with a $10\%$ energy spread. Energy spread control is under study through beam loading and bunch profile manipulation. A ``fish-bone'' structure starts forming in the driver bunch and both bunches are transversely focused where there is no significant bunch length change. 2D field distribution for the two-beam case is given in Fig.\ref{fig:2D_field} with accelerating blue region and decelerating red region. 
\begin{figure}[htb!] 
\centering
\subfloat[]{\includegraphics[width=0.45\textwidth] {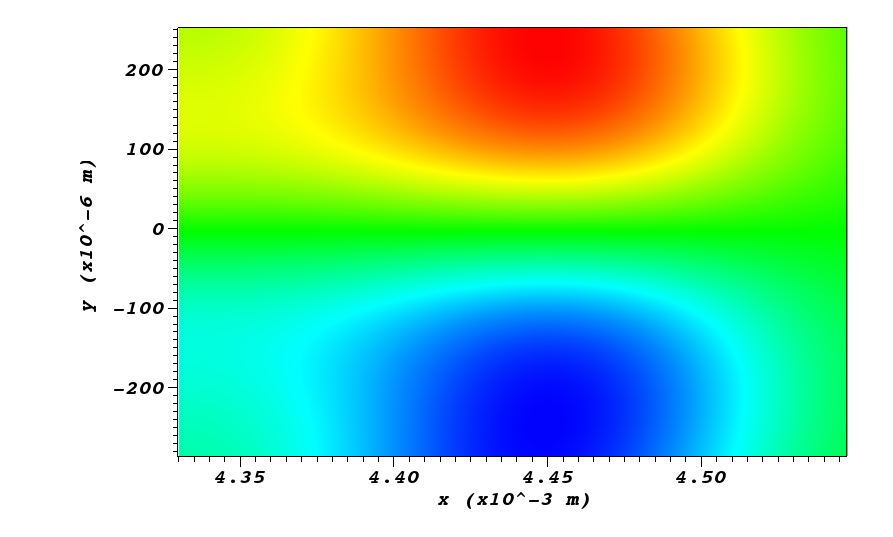}} \\
\subfloat[]{\includegraphics[width=0.45\textwidth] {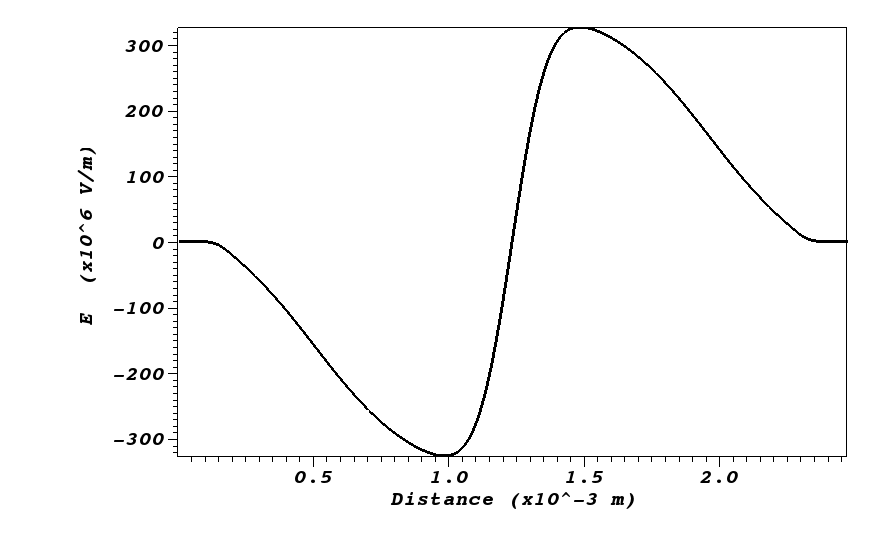}} 
\caption{(a) 2D and (b) projected field strength of the transverse plasma wakefield.}
\label{fig:foc_field}
\vspace{-1.5em}
\end{figure}
\begin{figure}[htb] 
\raggedleft
\includegraphics[width=0.450\textwidth] {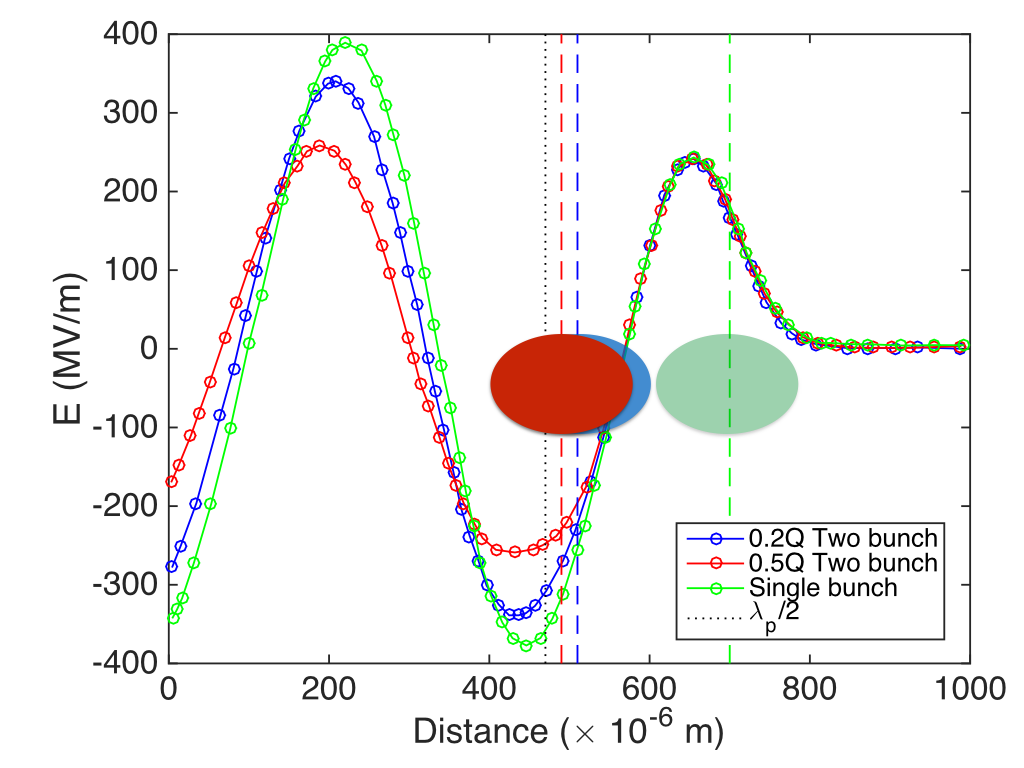}
\caption{Longitudinal instantaneous wakefield components in single and two-bunch cases after $0.5\,$m propagation with the hint of beam loading. Oval shapes represent bunches of different species on their respective locations with the same colour code as the curves.}
\label{fig:beam_loading}
\vspace{-1.5em}
\end{figure}
Plasma wakefields consist of co-existing transverse and longitudinal fields as they are induced due to the motion of the plasma electrons in both directions. The transverse field accompanying the above longitudinal field is shown in Fig.\ref{fig:foc_field}-(a) as a 2D intensity map. The transverse fields generally have a comparable field strength to the longitudinal fields (Fig.\ref{fig:foc_field}-(b)) and they might act as focusing fields depending on the phase of the trailing bunch.
\section{Beam loading with two bunches} 
It has been observed that in the presence of a second bunch the resulting field in the region is modified by the self electromagnetic field of this bunch. This phenomena is similar to the ``beam loading'' in the RF cavities. The position of the second beam can be adjusted so that the beam loading effect can be used, in favour of the scheme, in a way to reduce the energy spread on the second bunch by adjusting the field gradient across it. Such a scan was performed by moving the trailing bunch around its initial location at $\lambda_p/2$. Effect of the beam loading on controlling the energy spread is shown in Fig.\ref{fig:energy_distance} previously. This phenomenon was demonstrated in Fig.\ref{fig:beam_loading} in more detail for the trailing bunch locations providing the highest final energies for each case. Longitudinal plasma wakefield components in the presence of a single bunch and two bunch cases with different trailing beam charges are presented. 

As seen in the figure, the absolute values and the shape of the first bucket located about $450\,\mu$m are modified for different charges of the trailing bunch. Beam loading is larger for a trailing bunch carrying $50\%$ of the drive bunch and the a plateau is formed leading to a $1.4\%$ smaller energy spread compared to the case where the trailing bunch has $20\%$ of the drive beam charge. The cause of the modification is both the charge difference and the fact that the locations of trailing bunches are $20\,\mu$m apart from each other for presented cases. This effect is under further optimisation.

\section{Diagnostics under consideration} 
The novelty of the experimental study of plasma wakefields brings the necessity of employing and developing the novel measurement techniques for both plasma and the beam.

In PARS experimental station we aim to implement various measurement techniques such as optical transmission radiation interference (OTRI) technique \cite{OTRI}, coherent diffraction radiation (CDR) monitoring \cite{CDR1,CDR2,CDR3}, electro-optical sampling (EOS) \cite{EOS} etc. for the electron beams and plasma diagnostics to measure beam sizes and densities. A double-magnet magnetic spectrometer with a segmented beam dump is being studied to demonstrate the feasibility to measure a wide range of energies with an adequate energy resolution \cite{segdump}.
\section{Conclusions and outlook} 
 A summary of feasibility studies for an electron driven plasma wakefield acceleration test facility is given in this paper. Given the realistic beam parameters, primarily an acceleration gradient of $200-300\,$MV/m is aimed to be experimentally demonstrated.
 
 The beam loading effect was studied and promising results were obtained towards the control of the energy spread of the trailing bunch. A $1.4\%$ improvement is shown in this paper, further studies considering different drive and trailing bunch profiles are in progress in order to optimise the transformer ratio and energy spread.

Alongside with plasma acceleration studies, plasma lensing effect will be tested as well, which is reported elsewhere \cite{plasma_lens}.
\section{Acknowledgements}
This work was supported by the Cockcroft Institute Core Grant and STFC.

The authors gratefully acknowledge the computing time granted on the supercomputer JUROPA at J{\"u}lich Supercomputing Centre (JSC).
%
\raggedend


\begin{thebibliography}{1}   
\bibitem{Clara} J.A Clarke, et al.,  JINST 9 T05001 (2014).
\bibitem{Xia} G. Xia et al., NIMA 740 165-172 (2014).
\bibitem{VSim}  C. Nieter, and J. R. Cary, Journal of Computational Physics, vol. 196, pp. 448-473 (2004). 
\bibitem{OTRI} R. B. Fiorito et al., PRSTAB 9, 052802 (2006). 
\bibitem{CDR1} A.H. Lumpkin, N.S. Sereno, D.W. Rule, Nucl. Instr. and Meth. A 475, pp. 470-475  (2001).
\bibitem{CDR2} M. Castellano, et al., Phys. Rev. E 63, 056501 (2001).
\bibitem{CDR3} B. Feng, et al., Nucl. Instr. and Meth. A 475, pp. 492-497 (2001).
\bibitem{EOS} T. Srinivasan-Rao et al., PRSTAB 5, 042801 (2002).
\bibitem{segdump} T. Lefevre et al., Proceedings of IPAC2007, WEPC14 (2007).  
\bibitem{plasma_lens} K. Hanahoe, these proceedings (2015).
\end{thebibliography}
\end{document}